\documentclass[fleqn,12pt,twoside]{article}
\usepackage{espcrc1}

\usepackage{graphicx}
\usepackage[figuresright]{rotating}
\usepackage{amssymb}

\newcommand{\AUL}{\ensuremath{A_\mathrm{UL}} }
\newcommand{\AULp}{\ensuremath{A_\mathrm{UL}(\phi)} }
\newcommand{\AULsp}{\ensuremath{A_\mathrm{UL}^{\sin\phi}} }
\newcommand{\AULstwop}{\ensuremath{A_\mathrm{UL}^{\sin 2\phi}} }

\newcommand{\pip}{\ensuremath{\pi^+} }
\newcommand{\pim}{\ensuremath{\pi^-} }
\newcommand{\piz}{\ensuremath{\pi^0} }
\newcommand{\XQSM}{$\chi$QSM}
\newcommand{\HDratio}{\ensuremath{|H_1^\perp / D_1|} }
\graphicspath{{/home/gnome/doc/hermes/plots/}
	{/home/gnome/doc/hermes/plots/figs/}
	{/home/gnome/doc/hermes/plots/world/}
	{/home/gnome/doc/hermes/plots/predict/}}
\title{Transverse Spin: HERMES Results and Future Plans}

\author{N.C.R. Makins%
\address{University of Illinois, 
	1110 W Green St., Urbana, IL, USA  61801-3080},
	for the HERMES Collaboration}
       
\begin{document}

\maketitle

\begin{abstract}
  Results from the HERMES experiment are presented on single-spin
  asymmetries in semi-inclusive hadron production from
  longitudinally polarized targets. 
  The data are compared with a number of theoretical calculations
  which relate the azimuthal dependence of the
  asymmetries to the transversity structure function $h_1(x)$.
\end{abstract}


\section{TRANSVERSITY AND FRIENDS}

It has come to light that a complete description of the quark distributions
in the proton at leading twist requires not only the structure functions 
$f_1(x)$ and $g_1(x)$, but also the transversity distribution $h_1(x)$.
This new structure function represents 
the degree to which the quarks are polarized along the proton's spin
direction when the proton is polarized transversely to the virtual photon.
The unique properties of $h_1(x)$ were beautifully presented by R.~Jaffe 
at this conference. 
The majority of this report is devoted to recent data sensitive to $h_1(x)$.

\begin{figure}[t]
  \mbox{a)}\hspace*{1ex}
  \includegraphics[width=7cm]{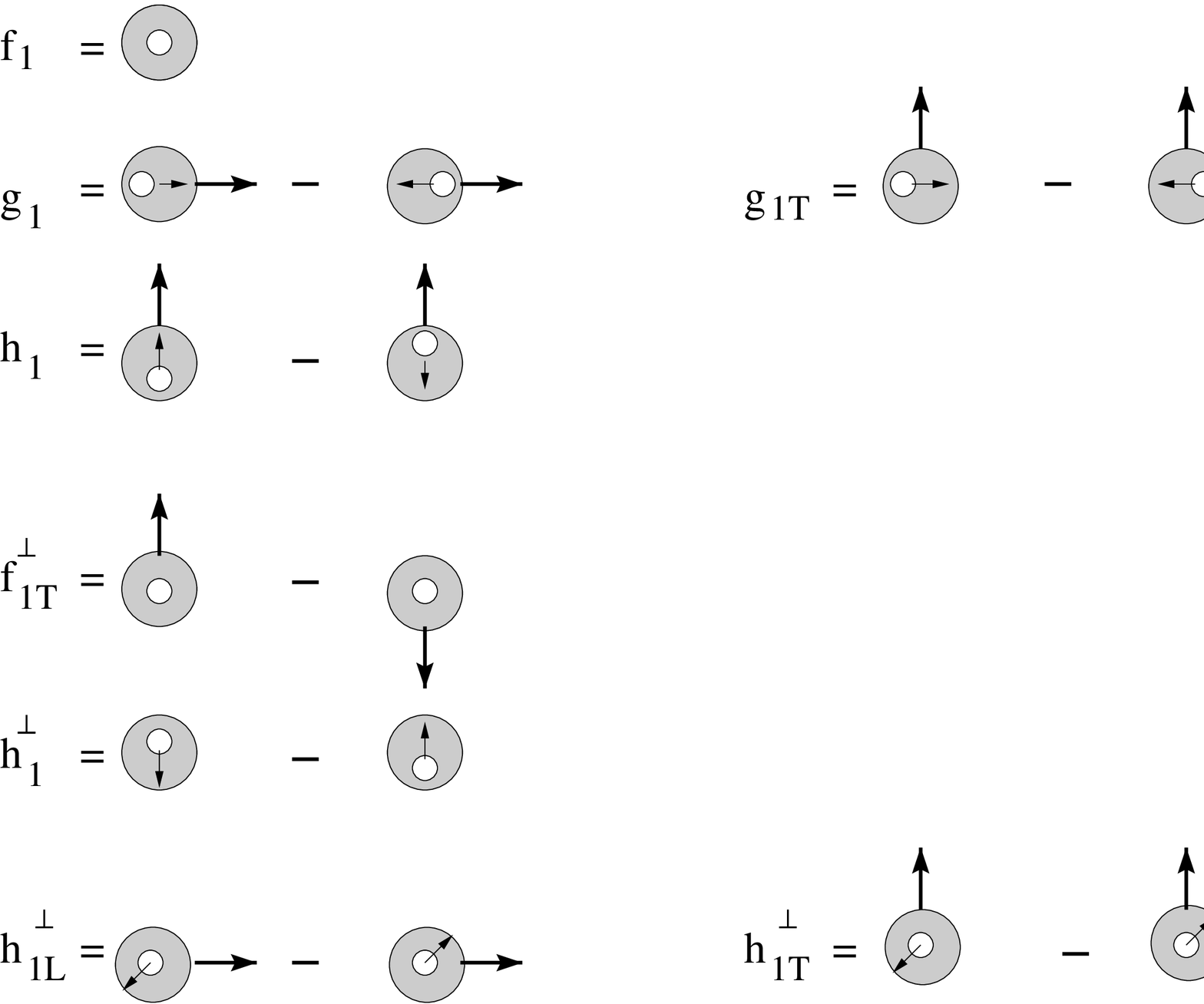}
  \hspace*{1.5em}\mbox{(b)}\hspace*{-2em}
  \raisebox{1em}{\includegraphics[width=8cm]{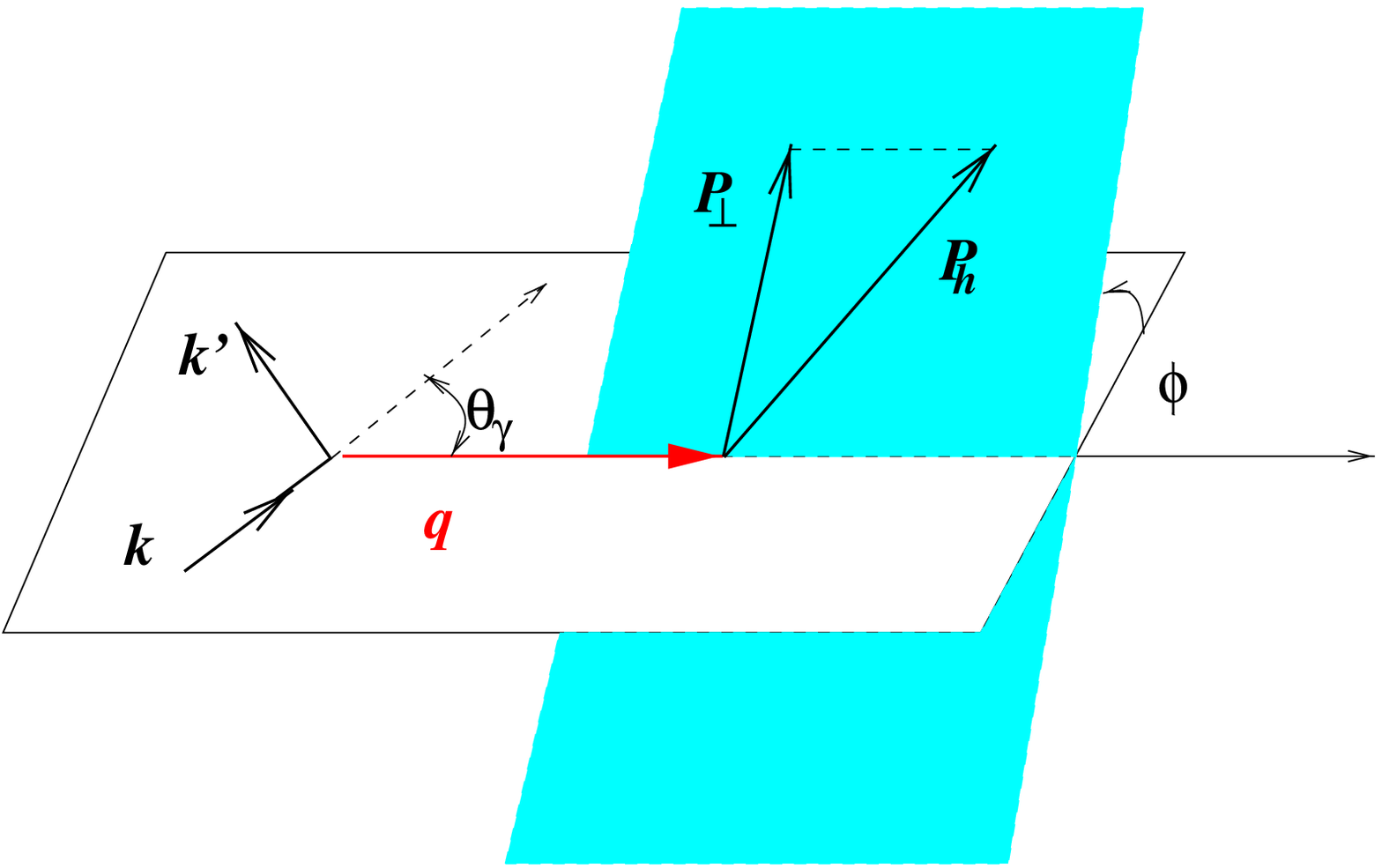}}
  \caption{
        (a) Summary of the classification scheme of~\protect\cite{Bible}
        for leading-twist distribution functions.
	(b) Diagram of SIDIS kinematics.}
  \label{fig:stoplights} 
\end{figure}

However, one may place transversity in a larger context.
In 1996, Mulders and Tangerman performed a complete tree-level analysis
of the semi-inclusive deep-inelastic scattering (SIDIS) cross-section,
taking into consideration all measurable spin-degrees of freedom in the 
initial and final state~\cite{Bible}.
Their work identified a series of 8 structure
functions of the proton at leading twist, along with an analogous set of 
8 fragmentation functions. These functions are illustrated graphically
in Fig.~\ref{fig:stoplights}a. In each picture, the virtual
photon probe is assumed to be incident from the left, the
large and small circles represent hadrons and quarks respectively,
and the arrows indicate their spin directions. The notation for the
fragmentation functions is obtained by replacing the letters $f$, $g$,
and $h$ with $D$, $G$, and $H$ respectively.

Each of these functions describes qualitatively different information
about hadronic structure and formation. 
The functions $f_1(x)$, $g_1(x)$, and $h_1(x)$ 
are the only ones which survive on integration over 
transverse momentum $k_T$. 
(The same is true of the analogous fragmentation functions.)
The other functions
are all implicitly dependent on intrinsic quark transverse motion,
which is necessarily related to the unknown orbital angular momentum
of quarks in the nucleon. 
New data are providing
first glimpses of several functions in this table, and much more 
can be expected in the near future.

The Mulders decomposition of the cross-section
reveals that experiments may access these new structure and fragmentation
functions by measuring \textit{azimuthal moments} in spin-dependent SIDIS.
Fig.~\ref{fig:stoplights}(b) illustrates the HERMES definition of
$\phi$, the azimuthal angle of
the measured final-state hadron, around the virtual photon direction, and
relative to the lepton scattering plane. 
Specifically one must perform an azimuthal decomposition of 
various cross-section terms $d\sigma_\mathrm{ABC}$, where A, B, and C
represent the spin-polarization of the lepton beam, proton target, 
and final-state hadron respectively. These subscripts take the values 
L, T, and U, denoting longitudinal, transverse, and no polarization.
The spin of the final-state hadron can usually be accessed only by
$\Lambda$-production measurements: the weak decay $\Lambda \rightarrow p \pim$
allows the hyperon's spin to be determined from the angular distribution 
of its decay products.


\section{THE HERMES EXPERIMENT}

The HERMES experiment has been taking data at the HERA accelerator 
in Hamburg, Germany since 1995. 
HERMES scatters longitudinally polarized electron and positron
beams of 27.6 GeV from polarized gas targets internal to the beam pipe.
Pure atomic H, D, and $^3$He have been used (as well as a variety
of unpolarized nuclear targets). 
Featuring polarized beams and targets,
and an open-geometry spectrometer with good particle identification (PID),
HERMES is well suited to a study of the spin-dependent 
azimuthal moments of the SIDIS cross-section. 
The PID capabilities of the experiment were significantly enhanced in
1998 when the threshold \v{C}erenkov detector (used to identify 
pions above a momentum of 4 GeV) was upgraded to a 
Ring Imaging system (RICH). This new detector
provides full separation between charged pions, kaons, and protons
over essentially the entire momentum range of the experiment.

In September 2000 HERMES completed its first phase of data taking,
using longitudinally-polarized targets. 
The 1998-2000 period was particularly successful, yielding a very large 
data set ($>$ 8 million DIS events) from polarized deuterium.
HERMES is now entering its second running
phase which will continue until 2006. A cornerstone of this
period will be measurements from transversely-polarized targets,
with the specific goal of exploring the transversity structure function.
Prospects for Run 2 were described in the talk of K.~Rith at this
conference. 


\section{HERMES MEASUREMENTS OF \AULp}

HERMES has measured the azimuthal distribution of charged~\cite{HERMEScharged}
and neutral~\cite{HERMESneutral}
pions in the scattering of ``unpolarized'' positrons from a 
longitudinally-polarized hydrogen target. (As the HERA beam is always 
polarized, an ``unpolarized'' sample was achieved by the helicity-balancing
of data collected with opposite beam polarization.)
The measured quantity is the following single-spin asymmetry (SSA):
\begin{equation}
  \AULp = \frac{1}{P_T}
	\frac{N^+(\phi) - N^-(\phi)}{N^+(\phi) + N^-(\phi)}.
\end{equation}
Here $P_T$ is the target polarization and $\phi$ is the azimuthal 
angle of the pion described above. 
$N^+$ and $N^-$ 
represent the pion yields in each of the two target spin orientations,
with the superscript indicating the helicity of the target 
in the reaction's center-of-mass frame
(\textit{i.e.} $N^+$ is collected when the target spin is antiparallel to 
the lepton beam momentum in the lab frame).
This choice is motivated by the familiar definition of the 
double-spin asymmetry $A_\mathrm{LL}$.
It is also important to note that the HERMES definition of $\phi$ 
is \textit{not} the
same as that of the ``Collins angle'' $\phi_C$
which appears in a number of publications. 

The measured asymmetries show a significant $\sin\phi$ moment 
in the case of \pip and \piz production, while no 
$\phi$-dependence is seen in \pim production.
The $\sin 2\phi$ moments of the asymmetries were also analyzed and found
to be consistent with zero in all cases. 
The mean $Q^2$ of these measurements is around 1.6 GeV$^2$.

\begin{figure}[thb]
  \includegraphics[width=0.85\textwidth]{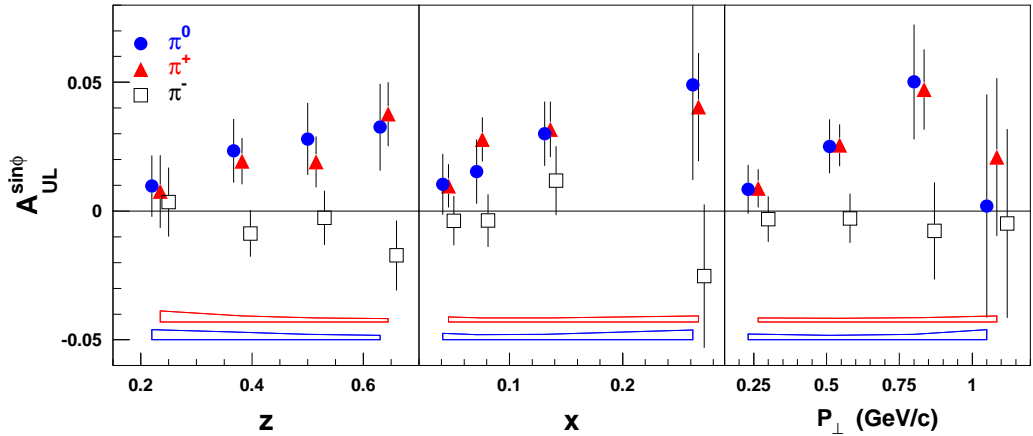}
  \vspace*{-1.5em}
  \caption{Kinematic dependences of \AULsp measured from a hydrogen 
	target~\protect\cite{HERMEScharged,HERMESneutral}.}
  \label{fig:azimkinH}
\end{figure}

At ``zeroth-order'' in complexity, the asymmetry moment \AULsp
(also termed analyzing power)
is related to the product of transversity $h_1(x)$ and the Collins
fragmentation function $H_1^\perp(z)$. 
This latter function provides a ``polarimeter'' for initial-state quark
polarization: it correlates the transverse spin of the struck 
quark with the angular distribution of hadrons in the jet it generates.
Fig.~\ref{fig:azimkinH} shows the dependence of 
\AULsp on the Bjorken scaling variable $x$, 
the energy fraction $z \equiv E_h/\nu$ of the pion, and its transverse
momentum $p_T$ relative to the virtual photon direction.
The dependences match the qualitative predictions of 
Collins~\cite{Collins} that the effect should
peak in the valence region $x \simeq 0.3$, rise with $p_T$ up to 
a maximum at some hadronic scale $0.3\,\textrm{--}\,0.9$ GeV, 
and be larger for \pip than \pim production. 

The reason for this last expectation can be shown by a simple 
calculation. It is reasonable to guess that the transverse quark polarization
($\delta q \equiv h_1^q$) is similar to the longitudinal quark polarization
($\Delta q \equiv g_1^q$) in the sense
$\delta d / \delta u \approx \Delta d / \Delta u \approx -1/2$. 
One may also estimate that the favoured and disfavoured
Collins fragmentation functions have a similar ratio as in the
unpolarized case 
($r \equiv D_d^{\pip} / D_u^{\pip} \approx (1-z)/(1+z) \approx 1/3$).
One then arrives at these simple estimates:
\begin{equation}
  A_p^{\pi^+} \approx \frac{4 \,\delta u + r \,\delta d}{4\,u + r\,d}
  \approx \frac{\delta u}{u},
	\hspace*{1.5em}
  A_p^{\pi^0} \approx \frac{4 \,\delta u + \,\delta d}{4\,u + d}
  \lesssim \frac{\delta u}{u},
	\hspace*{1.5em}
  A_p^{\pi^-} \approx \frac{4\,r \,\delta u + \,\delta d}{4\,r\,u + d}
  \approx 0
  \label{equ:envelopeH}
\end{equation}
The expectation of similar \pip and \piz asymmetries 
is also supported by the data.
	
\begin{figure}[thb]
  \hspace*{\fill}
  \includegraphics[width=0.31\textwidth]{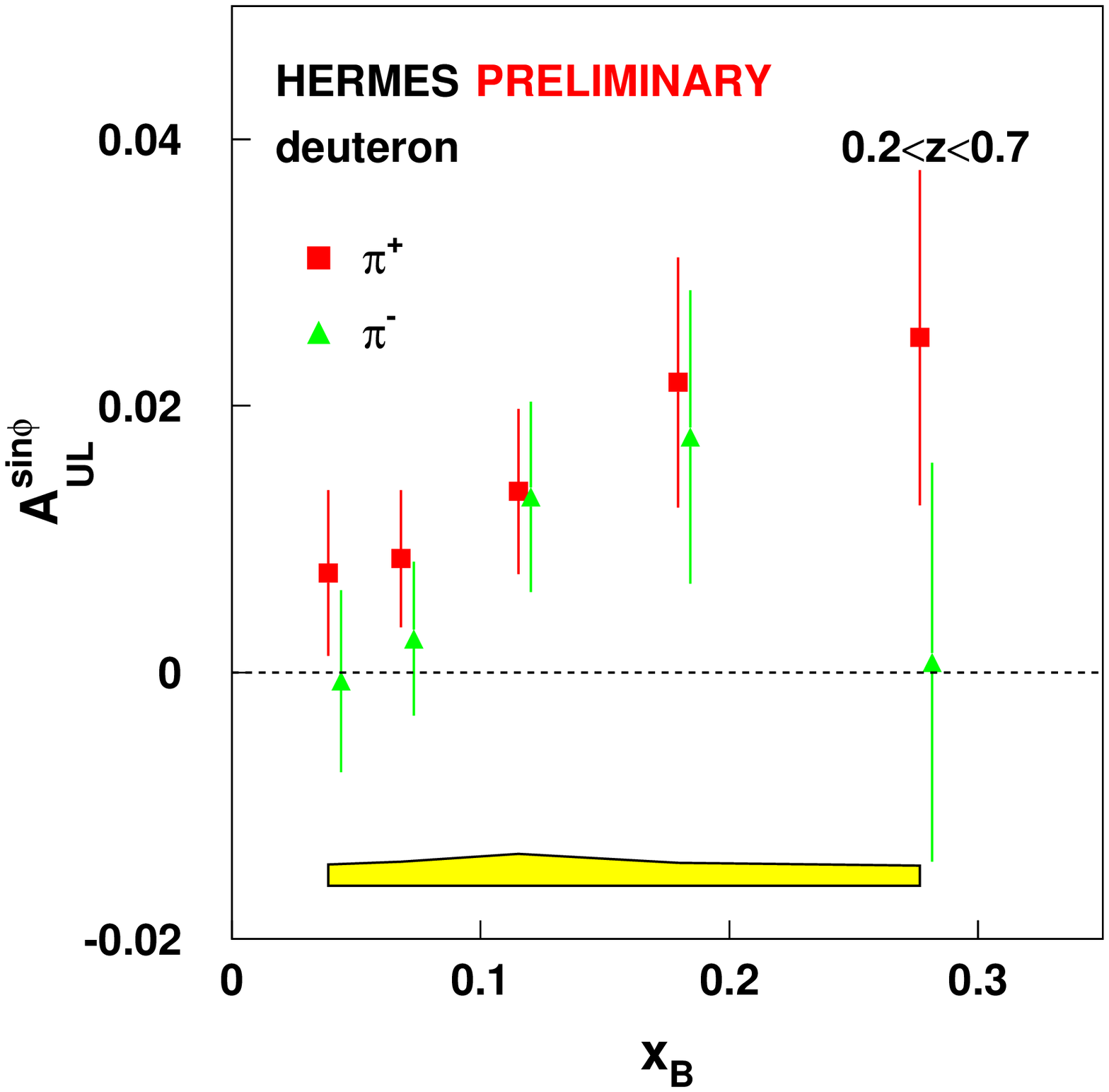}
  \hspace*{\fill}
  \includegraphics[width=0.31\textwidth]{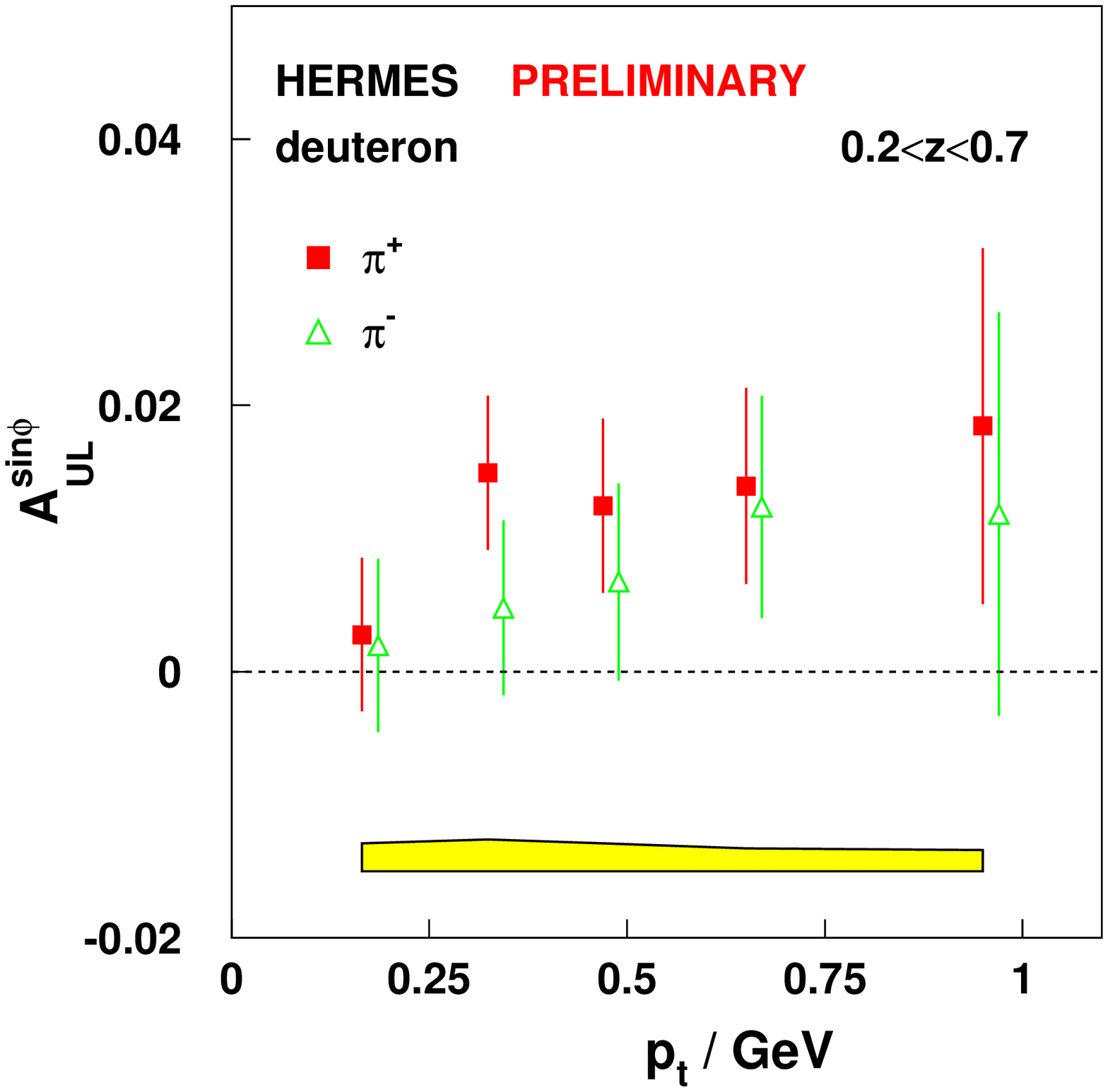}
  \hspace*{\fill}
  \includegraphics[width=0.31\textwidth]{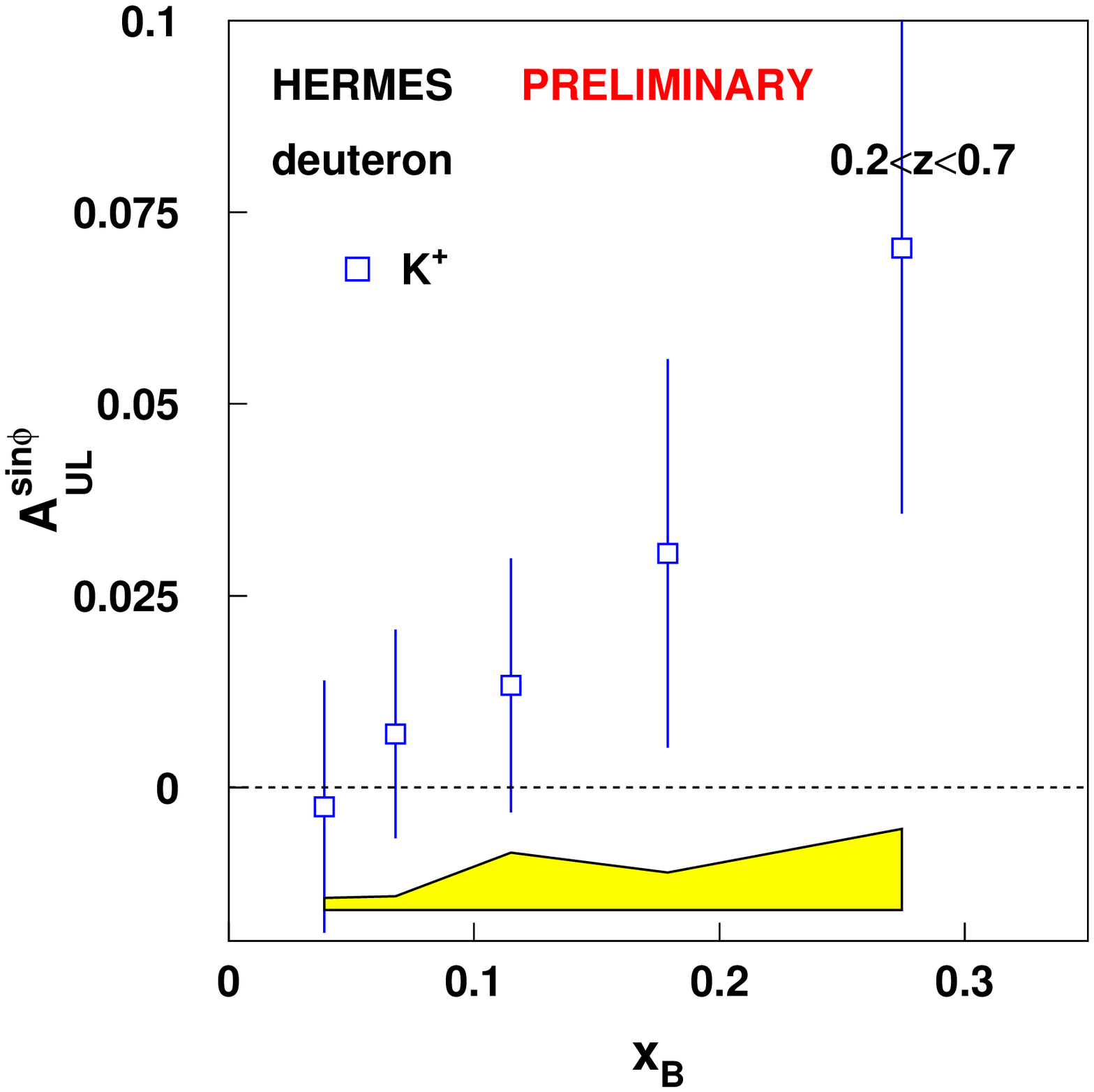}
  \hspace*{\fill}
  \vspace*{-1.5em}
  \caption{HERMES preliminary results on the analyzing power 
	\AULsp for charged pion and kaon production from a deuterium target.}
  \label{fig:azimkinD}
\end{figure}

HERMES has recently completed an analysis of \AUL
from the deuterium-target data collected in the years 1998 to 2000.
In Fig.~\ref{fig:azimkinD} preliminary results are presented for the 
analyzing power \AULsp for charged pion production.
A simple calculation of the type given in Eq.~\ref{equ:envelopeH} 
leads to the expectation that \AULsp from deuterium should be roughly 
the same for $\pi^+$, $\pi^-$, and $\pi^0$ production, 
and around half as large as for \pip production from hydrogen.
These qualitative expectations are indeed borne out by the data.
Spin-azimuthal asymmetries for charged kaon production have also been measured
for the first time, 
as shown in the right-hand panel of Fig.~\ref{fig:azimkinD}.

\subsection{Modelling \AULp}

For a more sophisticated interpretation of the data, one must consider
in detail which terms in the SIDIS cross-section of ref.~\cite{Bible}
contribute to the \AULp asymmetry. 
It is essential to realize that ``longitudinal'' and ``transverse''
target polarization have different meanings in experimental and theoretical
contexts. In experimental papers, a longitudinal target is polarized
along the \textit{lepton beam} direction, but in theoretical contexts
the relevant axis is the \textit{virtual photon}
direction. To distinguish the two, we use the 
notation $A_\mathrm{ABC}^w$ for moments $w$ of asymmetries in
the experimental convention, and the notation
$\langle w \rangle_\mathrm{ABC}$ of ref.~\cite{Bible}
to denote moments of the cross-section in the theoretical convention.
The HERMES measurement of \AULsp can thus be expressed as follows:
\begin{equation}
  \AULsp = \frac{
	S_L\,\langle \sin\phi \rangle_\mathrm{UL} + 
	S_T\,\langle \sin\phi \rangle_\mathrm{UT}}
	{S\,\langle 1 \rangle_\mathrm{UU}}.
\end{equation}
Here $S_L$ and $S_T$ represent the longitudinal and transverse projections 
of the target polarization $S$ onto the virtual photon direction.
At HERMES $S_L/S \approx 1$, and $S_T/S \approx 1/Q$.

The full theoretical decomposition of \AULsp and \AULstwop involves 
a number of functions at both leading and higher twist:
\[
  \langle 1 \rangle_\mathrm{UU} \sim f_1(x)\,D_1(z),
    \hspace*{1cm}
  \langle \sin\phi \rangle_\mathrm{UT} \sim 
	h_1(x)\,H_1^{\perp(1)}(z),
    \hspace*{1cm}
  \langle \sin 2\phi \rangle_\mathrm{UL} \sim 
	h_{1L}^{\perp(1)}(x)\,H_1^{\perp(1)}(z),
\]
\begin{equation}  
  \langle \sin\phi \rangle_\mathrm{UL} \sim \frac{1}{Q} [
	h_{1L}^{\perp(1)}(x)\,H_1^{\perp(1)}(z) \ \oplus\ 
	\tilde{h_L}(x)\,H_1^{\perp(1)}(z) \ \oplus\ 
	h_{1L}^{\perp(1)}(x)\,\tilde{H} ] 
  \label{eq:breakdown}
\end{equation}  
All kinematic prefactors have been suppressed in these expressions for brevity. 
The superscript $(1)$ appearing on several of the functions
denotes a $k_T^2$-weighted integral over transverse momentum.

The $\langle \sin\phi \rangle_\mathrm{UT}$ moment is directly proportional
to the product of transversity and the Collins function. However the 
present HERMES measurement is most directly related to
$\langle \sin\phi \rangle_\mathrm{UL}$, which is much more 
complex: it is sub-leading in $Q$, and contains the interaction-dependent
twist-3 functions $\tilde{h}_L$ and $\tilde{H}$. In addition 
the as-yet-unknown leading-twist distribution function 
$h_{1L}^\perp(x)$ makes an appearance (see Fig.~\ref{fig:stoplights}). 

It is necessary to make a few assumptions in order 
to proceed with modelling of the asymmetry.
The unknown functions $\tilde{h}_L$
and $h_{1L}^\perp$ are not unrelated; from Lorentz and rotational 
invariance one may derive the relation
\begin{equation}
  h_{1L}^{\perp(1)}(x) = x^2 \int_x^1 \frac{dy}{y^2} \left[
	-h_1(y) + \tilde{h}_L(y) + \frac{m_q}{M_p}\frac{g_1(y)}{y} \right] .
  \label{eq:Lorentz}
\end{equation}
Two approximations are commonly made in the literature.
The first is the ``reduced twist-3 approximation'',
which assumes that all twist-3 terms that are interaction-dependent 
or suppressed by the quark-mass $m_q$ are zero.
The second choice is to set the distribution
function $h_{1L}^\perp$ to zero. This last option is motivated by 
the HERMES measurement of $\AULstwop \approx 0$, as \AULstwop is proportional
to $h_{1L}^\perp$ (Eq.~\ref{eq:breakdown}).
With one of the functions $h_{1L}^\perp$ or $\tilde{h}_L$ set to zero,
the other can be calculated using Eq.~\ref{eq:Lorentz} and a model
for transversity.

The Collins fragmentation function is also unknown.
Its kinematic shape is typically taken from the 
heuristic parametrization of ref.~\cite{Collins},
\begin{equation}
  \frac{H_1^\perp(z,k_T)}{D_1(z)} = \eta \frac{M_C M_h |k_T|}{M_C^2 + k_T^2},
  \label{equ:collinsparam}
\end{equation}
with some normalization constant $\eta$ and hadronic scale 
$M_C \approx 2\,m_\pi\,\textrm{---}\,M_p$. 
These parameters may be constrained using hadron-production data
from DELPHI: a quark and antiquark produced from $Z^0$ decay
carry small but highly-correlated transverse polarizations. 
A pioneering analysis of these data~\cite{DELPHI} has yielded
the estimate $\HDratio = 6.3 \pm 1.7$~\%.
(Very recently, the value has been amended to 
$12.5 \pm 1.4$ \%~\cite{EfremovErratum}.)
Unfortunately this result is an average value integrated over
a rather poorly defined interval in $z$. Nevertheless, it provides an
important indication of the size of the Collins function, and one
which is independent of uncertainties in the distribution function sector.
As presented by A.~Ogawa at this workshop, high statistics data from 
the BELLE experiment will soon be analyzed 
in a similar fashion 
to yield much more precise results on $H_1^\perp$. 


\subsection{Comparison with Model Calculations}

A variety of theoretical calculations have been performed to address
the HERMES measurements of $A_\mathrm{UL}(\phi)$.

The work of ref.~\cite{Ogan00} attempts to explain the data with
a range of simple ans\"{a}tze. In Fig.~\ref{fig:ansatze}(a) and (b), the
solid curves correspond to the assumption that $h_{1L}^\perp = 0$,
while the dashed curves correspond to the ``reduced twist-3'' 
approximation $\tilde{h}_L = 0$. In each case, two guesses are made
for the magnitude of the transversity distribution: $h_1 = g_1$,
and saturation of the Soffer bound $h_1 = (f_1 + g_1)/2$.
The calculations show that the majority of the
asymmetry (around 75\%) 
originates from the subleading term $\langle \sin\phi \rangle_\mathrm{UL}$ 
associated with the longitudinal component of the target polarization, 
with only a quarter coming from the leading twist term 
$\langle \sin\phi \rangle_\mathrm{UT}$. This fact is reflected in
the curves: \AULsp is more sensitive to the ansatz made for the
higher-twist functions (solid vs dashed curves) than to the magnitude
of $h_1$ itself. Also, the data appear to indicate that the 
interaction-dependent twist-3 function $\tilde{h}_L$ cannot be ignored,
while the unknown twist-2 function $h_{1L}^\perp$ is likely of small 
magnitude.

\newlength{\xxht} \setlength{\xxht}{1.65in}
\begin{figure}[thb]
  \hspace*{\fill}
  \mbox{a)}
  \includegraphics[bb=18 37 511 518,height=\xxht]{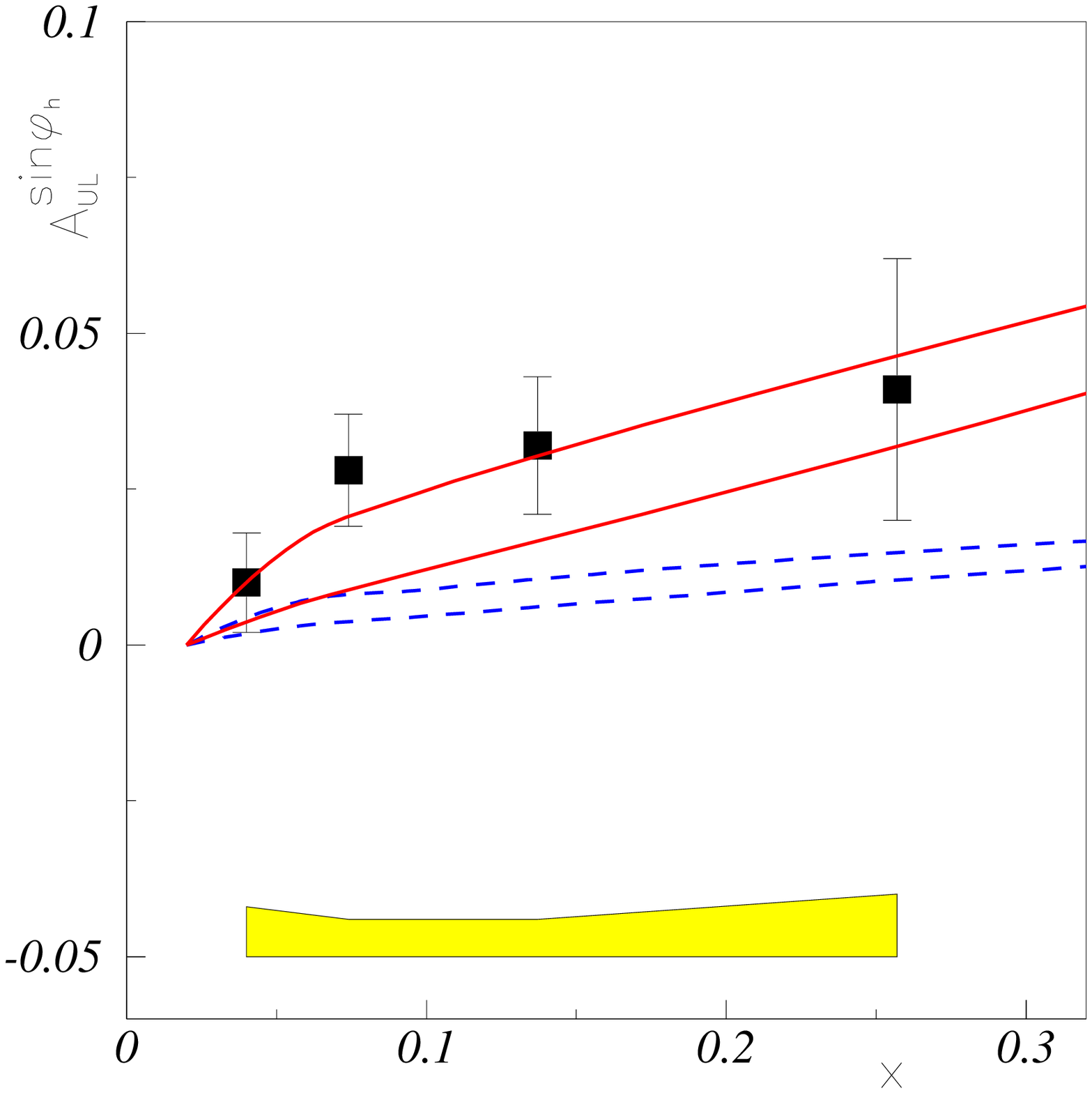}
  \hspace*{\fill}
  \mbox{b)}
  \includegraphics[bb=8 37 511 518,height=\xxht]{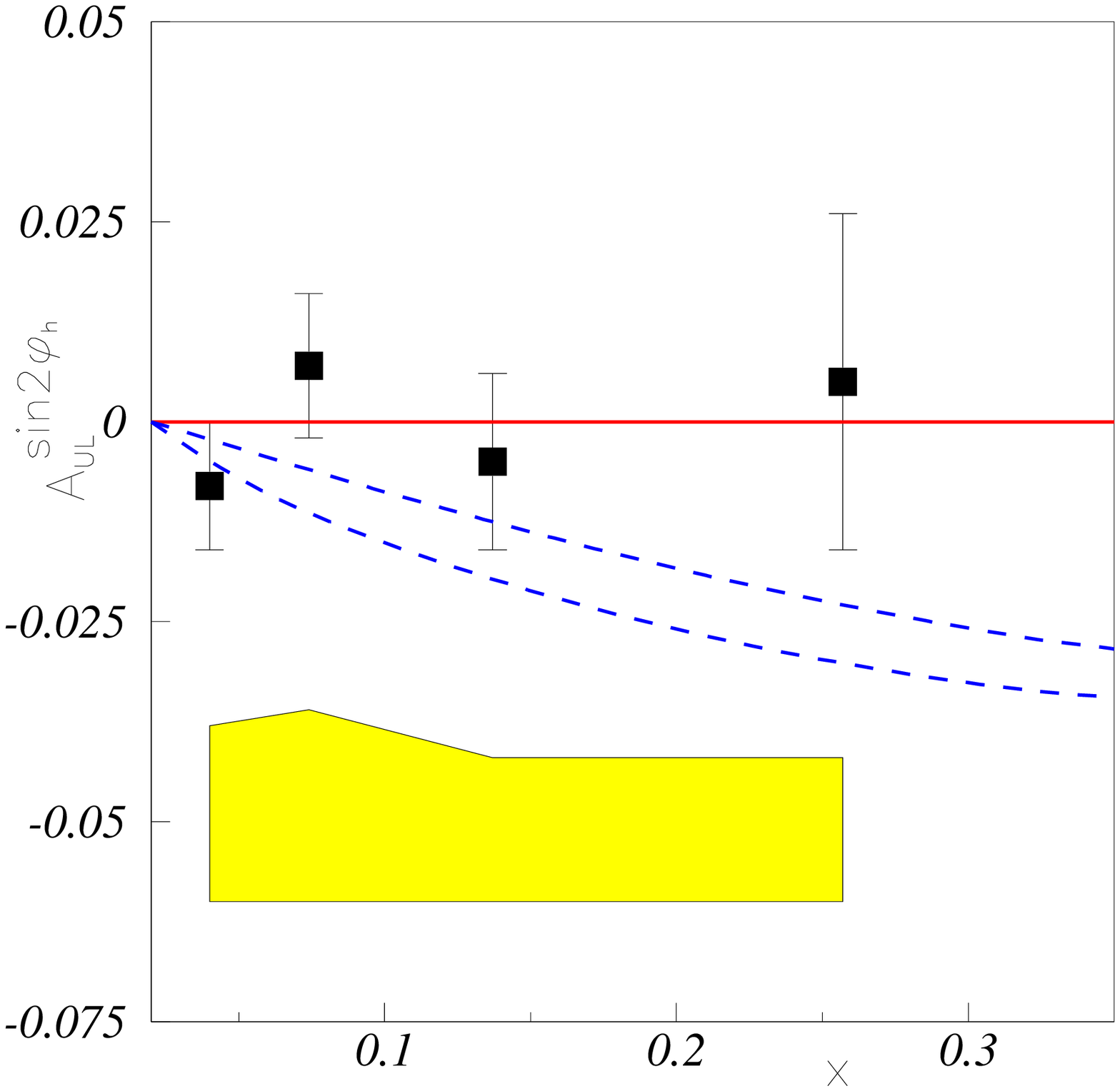}
  \hspace*{\fill}
  \mbox{c)}\hspace*{-1.5em}
  \includegraphics[clip,bb=95 65 431 340,height=\xxht]{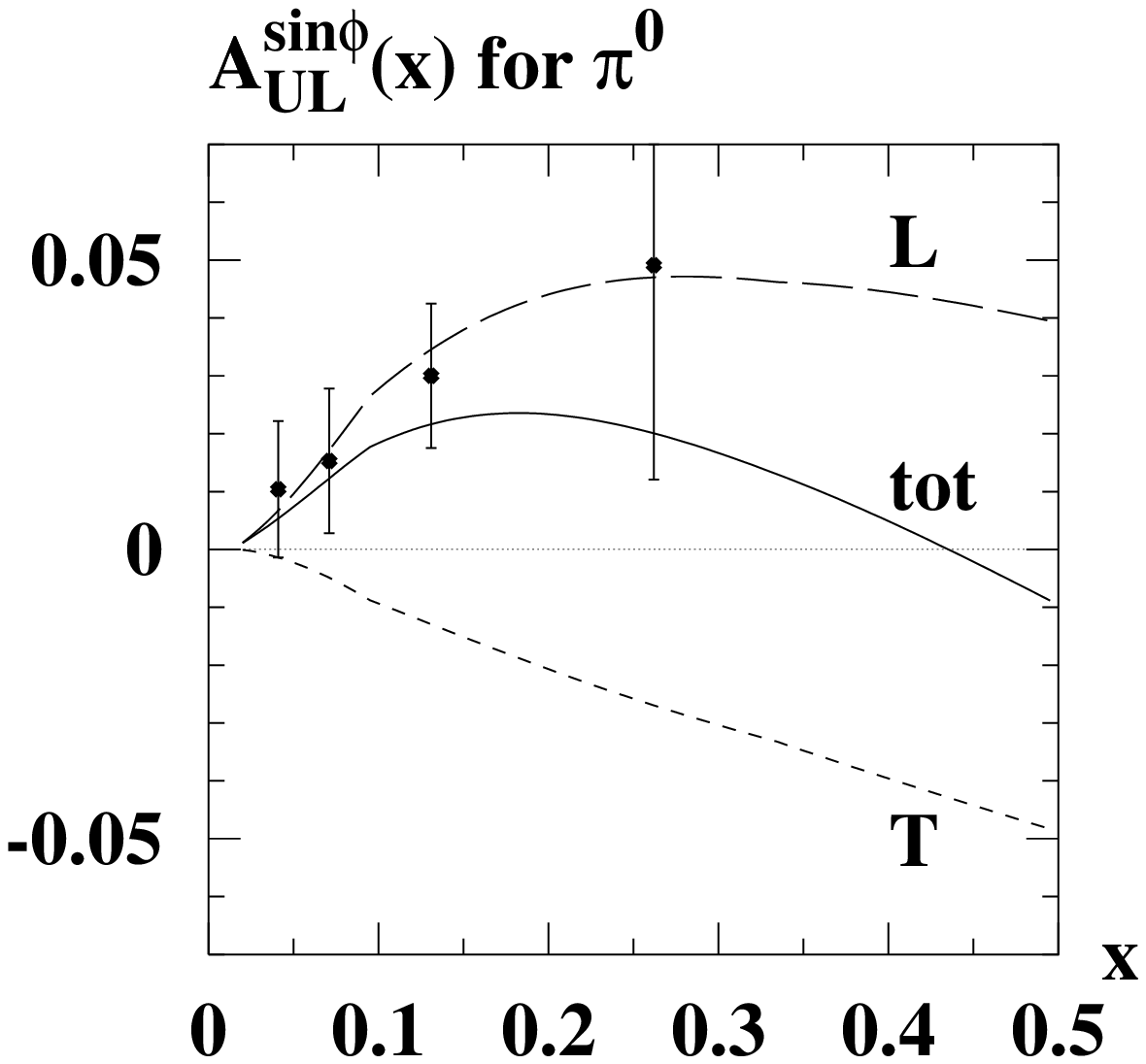}
  \hspace*{\fill}
  \vspace*{-1em}
  \caption{Calculations from ref.~\protect\cite{Ogan00} 
	of (a) \AULsp and (b) \AULstwop for \pip production from 
	a hydrogen target. Panel (c) shows the \XQSM~calculation from
	ref.~\protect\cite{EfremovErratum} of \AULsp for \piz production;
	the L and T curves show the contributions from the 
	longitudinal and transverse components of the target polarization.}
  \label{fig:ansatze}
\end{figure}

In ref.~\cite{EfremovPion}, $h_1(x)$ has been calculated directly in the 
Chiral-Quark Soliton Model (\XQSM) and used to estimate the asymmetry moments
\AULsp and \AULstwop in the reduced twist-3 approximation $\tilde{h}_L = 0$.
The calculations agree with the HERMES measurements for both charged 
and neutral pions, and also indicate that the longitudinal term
$\langle \sin\phi \rangle_\mathrm{UL}$ makes the dominant contribution
to the effect (see Fig.~\ref{fig:ansatze}(c) for \piz production, 
taken from the corrected analysis of ref.~\cite{EfremovErratum}).
A second \XQSM~calculation can be found
in ref.~\cite{Wakamatsu01}. In this work, \textit{all} distribution
functions involved in the asymmetries were calculated:
$h_1$, $\tilde{h}_L$, and $h_{1L}^\perp$. The higher-twist function
$\tilde{h}_L$ is found to be of significant magnitude, especially
in the region $x < 0.2$, while the 
unknown leading-twist distribution function $h_{1L}^\perp$ is 
predicted to be small but non-zero.
The calculations for \AULsp and \AULstwop are again in agreement 
with the HERMES data, within experimental accuracy.

Calculations in other models have been performed~\cite{Ma02} 
and obtain reasonable agreement with the measurements. 
All calculations agree that \AULsp is dominated
by higher-twist effects, due to the longitudinal target polarization.
For a better understanding of transversity, the next step is clear:
high precision SIDIS data on a transversely-polarized target are required.


\section{GLOBAL ANALYSIS OF SINGLE SPIN ASYMMETRY DATA}

Earlier data from the Fermilab E704 
experiment, from DELPHI, and from SMC at CERN are also potentially related to 
transversity. It is worthwhile to consider whether
a consistent picture has emerged from theoretical analyses of these
data sets.


\subsection{E704 and the Sivers Function}

About 10 years ago, the Fermilab E704 experiment measured a large 
analyzing power $A_N$ in the inclusive production of pions from a transversely
polarized proton beam of 200 GeV and an unpolarized target~\cite{E704}. 
Unexpectedly, positive and neutral pions displayed a pronounced tendency
to be produced to ``beam-left'' (when looking downstream with the
beam polarization pointing upwards). Negative pions showed a similar
analyzing power, but in the opposite ``beam-right'' direction.

The observables $A_N$ and \AULsp are odd under the
application of naive time reversal, and must
arise from the non-perturbative part of the 
cross-section. In the factorization picture of ref.~\cite{Bible},
either a T-odd fragmentation function or a T-odd
distribution function must play a role. 
In the first case, the E704 analyzing power is sensitive to 
the product of transversity $h_1$ and the T-odd Collins fragmentation 
function $H_1^\perp$. The second possibility involves the
unknown T-odd distribution function $f_{1T}^\perp(x,k_T)$ first postulated
by Sivers~\cite{Sivers}, together with the familiar unpolarized
fragmentation function $D_1$. 
Fits performed in the 
Collins-only~\cite{dualfits-Collins} and Sivers-only~\cite{dualfits-Sivers}
scenarios demonstrate that the E704 data may be described with equal success by
either picture.

Considerable discussion occurred at this workshop concerning the 
nature of the Sivers function $f_{1T}^\perp$. 
Given the T-even nature of the strong and electromagnetic interactions,
any T-odd function must involve an interference of 
amplitudes~\cite{Todd}. The most obvious way to generate such an interference
is via inital- or final-state interactions. Both are possible in 
a hadronic-beam experiment such as E704. In deep-inelastic scattering
with lepton beams, initial state interactions (and so the Sivers mechanism)
should be greatly suppressed. It may thus be said that the HERMES measurement 
of \AUL provides the first conclusive evidence that the Collins function
and transversity are both non-zero and of significant magnitude.

However, other suggestions exist for the origin of $f_{1T}^\perp$,
including contributions from gluonic poles~\cite{QiuSterman}
and spin-isospin interactions~\cite{Drago}. Further experimental and 
theoretical work is needed to determine the magnitude of
this function, whether it plays any role in lepton DIS, and whether 
it exists at all at leading twist.
As with all other distribution functions that vanish on integration
over transverse momentum, the Sivers function 
must be related at some level to parton orbital motion. A tantalizing
physical interpretation of this function in the context of the E704
asymmetry may be found in ref.~\cite{ChouYang}.

Turning briefly to the fragmentation aspect of these measurements,
the existence of the Collins function is deeply interesting in its own right. 
As it must arise from some
interference mechanism, it teaches us that the fragmentation process
possesses a large degree of \textit{phase coherence}. 
Older data on transverse hyperon polarization from 
high-energy unpolarized experiments have already provided evidence
that this is the case~\cite{Hyperon}. They are sensitive to the 
``polarizing'' fragmentation function $D_{1T}^\perp(z,k_T)$ which is
also T-odd. It is rather surprising that such interference effects
persist at high energies, given the large number of amplitudes that must 
be involved in inclusive or semi-inclusive hadron production.


\subsection{Global Analysis of Single-Spin Asymmetries}

A tentative, qualitative concensus has begun to emerge among the 
various analyses of single-spin asymmetry data at high energies.
The available data from HERMES, E704, and SMC may all be described by 
the transversity distribution calculated in the \XQSM, a Sivers function 
of zero, and a Collins function of approximate magnitude 
$\HDratio \approx$ 7\%~\cite{EfremovPion,dualfits-Collins}.
This apparent concensus is qualitative at best: the uncertainties
on the present experimental data and the rather poorly defined $z$-ranges 
over which the data are integrated certainly leave room for the 
existence of the Sivers function, for example. Also, a pair of sign 
errors was recently uncovered in the analyses of the HERMES and DELPHI
data~\cite{EfremovErratum}. The updated estimates for the Collins function are 
larger than before: $\HDratio = 12.5 \pm 1.4$~\% from the DELPHI data and
$\HDratio = 13.8 \pm 2.8$~\% from the HERMES data. 
It is presently unclear to what degree such sign errors affect other
theoretical analyses of transversity-related measurements.


\section{FUTURE MEASUREMENTS OF TRANSVERSITY}

It is clear from Eq.~\ref{eq:breakdown} that the most direct access
to transversity in SIDIS lies in measurements with transversely 
polarized targets. 
A precise measurement of $A_\mathrm{UT}(\phi)$,
sensitive at leading twist to the product of $h_1(x)$ and 
$H_1^\perp(z)$, will form a cornerstone of HERMES Run 2. 
Besides $A_\mathrm{UT}(\phi)$,
HERMES has considered other observables that may be 
sensitive to $h_1$.
A promising candidate is the correlation between the transverse polarization
of the target and of final-state $\Lambda$ baryons,
described by the spin-transfer fragmentation function $H_1(z)$.
If this function is of significant
size, $\Lambda$ polarization can serve as a ``polarimeter'' for the quark
spin distribution in the target. However results from HERMES
indicate that the \textit{longitudinal} spin-transfer from struck quark
to $\Lambda$ ($G_1(z)$) is small in DIS at intermediate 
energies~\cite{HERMESDLL}.
This likely indicates that the transverse spin transfer will
also be small and that $\Lambda$ polarization will not serve as a 
useful tactic for future transversity measurements.

It is important to note that the next round of single-spin asymmetry
measurements \textit{will} be able to distinguish between the Collins
and Sivers mechanisms. The SIDIS cross-section term $d\sigma_\mathrm{UT}$
has two contributions according to ref.~\cite{Bible}: 
a $\sin(\phi_h^l + \phi_S^l)$ moment proportional to $h_1 \, H_1^\perp$,
and a $\sin(\phi_h^l - \phi_S^l)$ moment proportional to $f_{1T}^\perp \, D_1$. 
The symbols $\phi_S^l$ and $\phi_h^l$ indicate the angle of the target spin
and final-state hadron momentum respectively, with respect 
to the lepton scattering plane.
The angular dependence of the Collins term may be understood by
realizing that the spin directions of the struck quark ($\phi_q^l$)
and final-state quark ($\phi_{q'}^l$) are related by
$\phi_q^l = \pi - \phi_{q'}^l$~\cite{Collins}: 
when the virtual
photon is absorbed by the struck quark, the component of the quark's spin
parallel to the lepton scattering plane is flipped,
while the perpendicular component is left unaffected.
The Collins fragmentation function, which correlates hadron direction 
with the spin of the primary quark, produces a 
$\sin(\phi_h^l - \phi_{q'}^l)$ behavior. Transversity
relates $\phi_q^l$ to $\phi_S^l$, giving a net $\sin(\phi_h^l + \phi_S^l)$
dependence in the cross section. The Sivers function, on the other
hand, correlates quark transverse (orbital) motion with target spin.
This motion is transferred directly to the final-state hadrons by $D_1$,
giving a net $\sin(\phi_h^l - \phi_S^l)$ behavior.

Current HERMES data cannot distinguish between the two: $\phi_S^l$
is always zero or $\pi$ for a target polarized longitudinally in the 
laboratory frame.
Similarly, the E704 data could not distinguish between the moments
as their experimental apparatus did not permit a measurement of the jet axis 
around which the pions were produced.
But forthcoming HERMES data with a transversely-polarized target
will permit the independent variation of
the target spin angle $\phi_S^l$ and the hadron angle $\phi_h^l$. 

In conclusion, data collected in recent years have provided a first glimpse of 
the transversity distribution and the Collins fragmentation function.
The data are still of modest precision, but they ``make sense'': 
the measurements can be described by a variety of theoretical models
and resonable assumptions. 
This baseline understanding has provided clear guidance
on which measurements to perform next. The upcoming round of results
from HERMES, COMPASS, RHIC-spin, and BELLE will provide much more precise
information on transversity and on other new structure and fragmentation 
functions.


\end{document}